\documentclass[twocolumn,tight]{aastex631}
\usepackage[T1]{fontenc}

\usepackage{graphicx}
\usepackage{hyperref}

\usepackage{gensymb}
\usepackage{amssymb}
\usepackage{booktabs}
\usepackage{amsmath}
\usepackage{enumerate}
\newcommand{\farcsec}{\mbox{\ensuremath{\hspace{1pt}.\!\!^{\prime\prime}}}}%

\renewcommand*{\figureautorefname}{Fig.}

\renewcommand*{\sectionautorefname}{Sect.}

\begin{document}

\title{exoALMA XVII. Characterizing the Gas Dynamics around Dust Asymmetries}

\correspondingauthor{Lisa Wölfer}
\email{lwoelfer@mit.edu}

\author[0000-0002-7212-2416]{Lisa Wölfer}
\affiliation{Department of Earth, Atmospheric, and Planetary Sciences, Massachusetts Institute of Technology, Cambridge, MA 02139, USA}

\author[0000-0001-6378-7873]{Marcelo Barraza-Alfaro}
\affiliation{Department of Earth, Atmospheric, and Planetary Sciences, Massachusetts Institute of Technology, Cambridge, MA 02139, USA}

\author[0000-0003-1534-5186]{Richard Teague}
\affiliation{Department of Earth, Atmospheric, and Planetary Sciences, Massachusetts Institute of Technology, Cambridge, MA 02139, USA}

\author[0000-0003-2045-2154]{Pietro Curone} 
\affiliation{Dipartimento di Fisica, Universit\`a degli Studi di Milano, Via Celoria 16, 20133 Milano, Italy}
\affiliation{Departamento de Astronom\'ia, Universidad de Chile, Camino El Observatorio 1515, Las Condes, Santiago, Chile}

\author[0000-0002-7695-7605]{Myriam Benisty}
\affiliation{Universit\'e C\^ote d'Azur, Observatoire de la C\^ote d'Azur, CNRS, Laboratoire Lagrange, France}
\affiliation{Max-Planck Institute for Astronomy (MPIA), Königstuhl 17, 69117 Heidelberg, Germany}

\author[0000-0003-1117-9213]{Misato Fukagawa} 
\affiliation{National Astronomical Observatory of Japan, 2-21-1 Osawa, Mitaka, Tokyo 181-8588, Japan}


\author[0000-0001-7258-770X]{Jaehan Bae}
\affiliation{Department of Astronomy, University of Florida, Gainesville, FL 32611, USA}



\author[0000-0002-2700-9676]{Gianni Cataldi} 
\affiliation{National Astronomical Observatory of Japan, 2-21-1 Osawa, Mitaka, Tokyo 181-8588, Japan}

\author[0000-0002-1483-8811]{Ian Czekala}
\affiliation{School of Physics \& Astronomy, University of St. Andrews, North Haugh, St. Andrews KY16 9SS, UK}


\author[0000-0003-4689-2684]{Stefano Facchini}
\affiliation{Dipartimento di Fisica, Universit\`a degli Studi di Milano, Via Celoria 16, 20133 Milano, Italy}

\author[0000-0003-4679-4072]{Daniele Fasano} 
\affiliation{Universit\'e C\^ote d'Azur, Observatoire de la C\^ote d'Azur, CNRS, Laboratoire Lagrange, France}

\author[0000-0002-9298-3029]{Mario Flock} 
\affiliation{Max-Planck Institute for Astronomy (MPIA), Königstuhl 17, 69117 Heidelberg, Germany}


\author[0000-0002-5503-5476]{Maria Galloway-Sprietsma}
\affiliation{Department of Astronomy, University of Florida, Gainesville, FL 32611, USA}

\author[0000-0002-5910-4598]{Himanshi Garg}
\affiliation{School of Physics and Astronomy, Monash University, VIC 3800, Australia}

\author[0000-0002-8138-0425]{Cassandra Hall} 
\affiliation{Department of Physics and Astronomy, The University of Georgia, Athens, GA 30602, USA}
\affiliation{Center for Simulational Physics, The University of Georgia, Athens, GA 30602, USA}
\affiliation{Institute for Artificial Intelligence, The University of Georgia, Athens, GA, 30602, USA}

\author[0000-0001-6947-6072]{Jane Huang} 
\affiliation{Department of Astronomy, Columbia University, 538 W. 120th Street, Pupin Hall, New York, NY, USA}

\author[0000-0003-1008-1142]{John~D.~Ilee} 
\affiliation{School of Physics and Astronomy, University of Leeds, Leeds, UK, LS2 9JT}

\author[0000-0001-8446-3026]{Andr\'es F. Izquierdo}
\altaffiliation{NASA Hubble Fellowship Program Sagan Fellow}
\affiliation{Department of Astronomy, University of Florida, Gainesville, FL 32611, USA}
\affiliation{Leiden Observatory, Leiden University, P.O. Box 9513, NL-2300 RA Leiden, The Netherlands}
\affiliation{European Southern Observatory, Karl-Schwarzschild-Str. 2, D-85748 Garching bei M\"unchen, Germany}

\author[0000-0001-7235-2417]{Kazuhiro Kanagawa} 
\affiliation{College of Science, Ibaraki University, 2-1-1 Bunkyo, Mito, Ibaraki 310-8512, Japan}

\author[0000-0002-8896-9435]{Geoffroy Lesur} 
\affiliation{Univ. Grenoble Alpes, CNRS, IPAG, 38000 Grenoble, France}

\author[0000-0003-4663-0318]{Cristiano Longarini} 
\affiliation{Institute of Astronomy, University of Cambridge, Madingley Road, CB3 0HA, Cambridge, UK}
\affiliation{Dipartimento di Fisica, Universit\`a degli Studi di Milano, Via Celoria 16, 20133 Milano, Italy}

\author[0000-0002-8932-1219]{Ryan A. Loomis}
\affiliation{National Radio Astronomy Observatory, 520 Edgemont Rd., Charlottesville, VA 22903, USA}


\author[0000-0002-1637-7393]{Francois Menard} 
\affiliation{Univ. Grenoble Alpes, CNRS, IPAG, 38000 Grenoble, France}

\author[0009-0002-1495-6321]{Anika Nath}
\affiliation{Department of Earth, Atmospheric, and Planetary Sciences, Massachusetts Institute of Technology, Cambridge, MA 02139, USA}

\author[0000-0003-4039-8933]{Ryuta Orihara} 
\affiliation{College of Science, Ibaraki University, 2-1-1 Bunkyo, Mito, Ibaraki 310-8512, Japan}

\author[0000-0001-5907-5179]{Christophe Pinte}
\affiliation{Univ. Grenoble Alpes, CNRS, IPAG, 38000 Grenoble, France}
\affiliation{School of Physics and Astronomy, Monash University, VIC 3800, Australia}

\author[0000-0002-4716-4235]{Daniel J. Price} 
\affiliation{School of Physics and Astronomy, Monash University, VIC 3800, Australia}

\author[0000-0003-4853-5736]{Giovanni Rosotti} 
\affiliation{Dipartimento di Fisica, Universit\`a degli Studi di Milano, Via Celoria 16, 20133 Milano, Italy}

\author[0000-0002-0491-143X]{Jochen Stadler} 
\affiliation{Universit\'e C\^ote d'Azur, Observatoire de la C\^ote d'Azur, CNRS, Laboratoire Lagrange, France}


\author[0000-0002-3468-9577]{Gaylor Wafflard-Fernandez} 
\affiliation{Univ. Grenoble Alpes, CNRS, IPAG, 38000 Grenoble, France}

\author[0000-0002-7501-9801]{Andrew J. Winter}
\affiliation{Universit\'e C\^ote d'Azur, Observatoire de la C\^ote d'Azur, CNRS, Laboratoire Lagrange, France}
\affiliation{Max-Planck Institute for Astronomy (MPIA), Königstuhl 17, 69117 Heidelberg, Germany}

\author[0000-0003-1412-893X]{Hsi-Wei Yen} 
\affiliation{Academia Sinica Institute of Astronomy \& Astrophysics, 11F of Astronomy-Mathematics Building, AS/NTU, No.1, Sec. 4, Roosevelt Rd, Taipei 10617, Taiwan}

\author[0000-0001-8002-8473	]{Tomohiro C. Yoshida} 
\affiliation{National Astronomical Observatory of Japan, 2-21-1 Osawa, Mitaka, Tokyo 181-8588, Japan}
\affiliation{Department of Astronomical Science, The Graduate University for Advanced Studies, SOKENDAI, 2-21-1 Osawa, Mitaka, Tokyo 181-8588, Japan}

\author[0000-0001-9319-1296	]{Brianna Zawadzki} 
\affiliation{Department of Astronomy, Van Vleck Observatory, Wesleyan University, 96 Foss Hill Drive, Middletown, CT 06459, USA}
\affiliation{Department of Astronomy \& Astrophysics, 525 Davey Laboratory, The Pennsylvania State University, University Park, PA 16802, USA}



\begin{abstract}
\noindent The key planet-formation processes in protoplanetary disks remain an active matter of research. One promising mechanism to radially and azimuthally trap millimeter-emitting dust grains, enabling them to concentrate and grow into planetesimals, is anticyclonic vortices. While dust observations have revealed crescent structures in several disks, observations of their kinematic signatures are still lacking. Studying the gas dynamics is, however, essential to confirm the presence of a vortex and understand its dust trapping properties. In this work, we make use of the high-resolution and sensitivity observations conducted by the exoALMA large program to search for such signatures in the $^{12}$CO and $^{13}$CO  molecular line emission of four disks with azimuthal dust asymmetries: HD\,135344B, HD\,143006, HD\,34282, and MWC\,758. To assess the vortex features, we constructed an analytical vortex model and performed hydrodynamical simulations. For the latter, we assumed two scenarios: a vortex triggered at the edge of a dead zone and of a gap created by a massive embedded planet. These models reveal a complex kinematical morphology of the vortex. When compared to the data, we find that none of the sources show a distinctive vortex signature around the dust crescents in the kinematics.    
%
HD\,135344B exhibits a prominent feature similar to the predictions from the simulations, thus making this the most promising target for sensitive follow-up studies at higher resolution and in particular with less abundant molecules at higher resolution and sensitivity, to trace closer to the disk midplane.
\end{abstract}
\section{Introduction}\label{sec:intro}
%
%

\noindent In recent years, high resolution and sensitivity observations of the dust and gas material in protoplanetary disks with facilities such as the Atacama Large Millimeter/submillimeter Array (ALMA; \citealp{ALMA2015}), the Spectro-Polarimetric High-contrast Exoplanet REsearch (SPHERE; \citealp{Sphere2019}), or the Gemini Planet Imager (GPI; \citealp{Macintosh2014}) have revealed that various substructures are ubiquitous in disks, and may indicate planet--disk interactions. In the dust, this includes spiral arms, gaps or cavities, rings, and azimuthal asymmetries (e.g., review in \citealp{Bae2023}). 
The latter two are associated with local pressure maxima, where dust grains can be radially and/or azimuthally trapped and prevented from undergoing processes such as inward drift or fragmentation (e.g., \citealp{Birnstiel2010b,Pinilla2012a,Pinilla2012b}). 
Dust traps can further enable the streaming instability \citep[e.g.,][]{Stammler2019}, thus allowing dust to grow to larger sizes and form planetesimals (e.g., \citealp{Bai2010a,Bai2010b,Raettig2015}). 

One possible origin of pressure bumps in disks is embedded planets, which - if massive enough - are capable of opening a gap in the disk and form symmetric dust rings (e.g., \citealp{Ayliffe2012,Dipierro2015,Rosotti2016,Dong2017,Dong2018,Dipierro2018}). In low viscosity regions, the edges of such gaps or cavities can then become unstable against the Rossby wave instability (RWI, \citealp{Lovelace1999,Li2000}), resulting in the formation of an anticyclonic vortex where dust can be trapped radially and azimuthally, possibly giving rise to the observed crescent-shaped asymmetries (e.g., \citealp{Birnstiel2013,Fu2014b,Bae2016}). 
Evidence of azimuthal dust trapping has been found in a number of sources with ALMA, seen as a large horseshoe structure (e.g. HD\,142527, \citealp{Casassus2013}; IRS\,48, \citealp{Marel2013}; AB\,Aur, \citealp{Tang2012,Tang2017}), or as a concentration of dust within a dust ring (e.g. HD\,135344B, SR\,21 \citealp{Perez2014}; V1247\,Ori \citealp{Kraus2017}; HD\,34282 \citealp{Plas2017}; MWC\,758 \citealp{Boehler2018}; HD\,143006, \citealp{perez2018}), with high resolution observations revealing more complex substructures in some of these sources (e.g. \citealp{Marel2016,Cazzoletti2018,Muro2020,Yang2023}).

Alternatively to dust trapping, azimuthal dust asymmetries may be caused in a circumbinary disk by a sufficiently massive companion carving an eccentric cavity, which results in a gas overdensity at the cavity edge and a horseshoe structure (e.g. \citealp{Ragusa2017,Price2018}). \cite{Marel2021} compare several disk observables in a number of asymmetric disks to distinguish between vortices, horseshoes created by binary companions, and spiral density waves, concluding that the current data do not allow for a clear classification. Apparent dust Asymmetries may further arise in inclined sources due to optically thick emission at the hot inner rim of a disk cavity \citep{Ribas2024,Guerra2024}.

The edge of a disk cavity or gap favors vortex formation regardless of the underlying process creating the structure, not limited to planets. The RWI can further be triggered by infalling material \citep{Bae2015,Kuznetsova2022} or at the edge of a dead-zone (e.g., \citealp{Flock2015,Flock2017a}). 
Moreover, there are other instabilities which can lead to the formation of vortices: the vertical shear instability \citep{Richard2016,Flock2020,Manger2020,Pfeil2021}, the convective overstability \citep{Lyra2014,Raettig2021}, or the zombie vortex instability (e.g., \citealp{Marcus2013,Marcus2015,Marcus2016}). 
To differentiate the mechanisms and understand the process of dust trapping in vortices it is important to study their gas dynamics. 
\begin{figure*}
\centering
\includegraphics[width=1.0\textwidth]{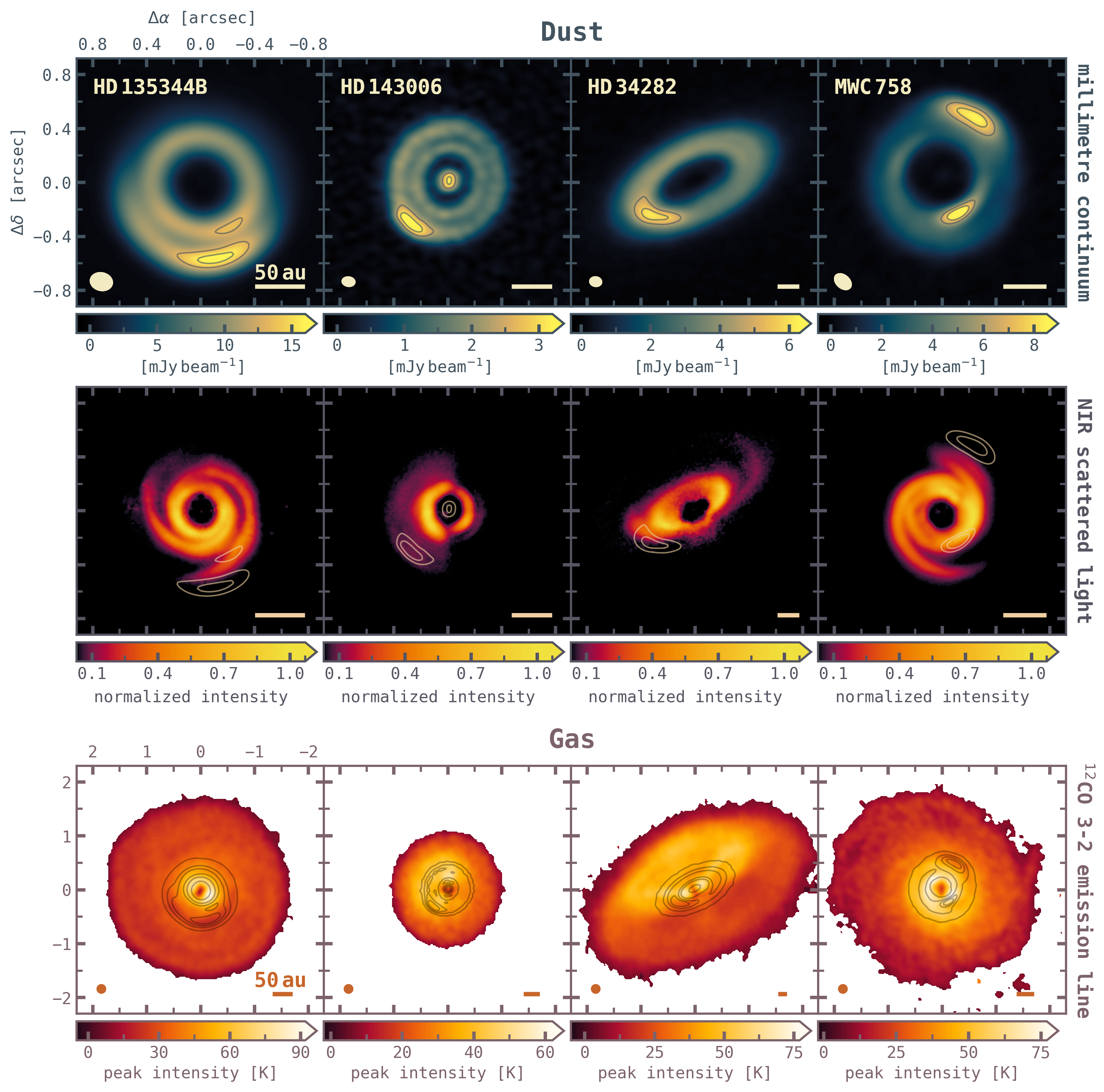}
\caption{Overview of the four sources studied in this work, shown on the same spatial scale for the dust (top two rows) and gas (bottom row). Top: Millimetre dust continuum as observed within the exoALMA program \citep{Curone_exoALMA}. Middle: Micron-sized dust observed through near-infrared scattered light with SPHERE \citep{Stolker2017,Boer2021,Ren2023}. The SPHERE images are normalized to the peak value and shown with a logarithmic normalization of the color map to emphasize the structures in the outer disk. To highlight the location of the crescents, two continuum contours around the peak of the dust asymmetries are overlaid on the dust images. Bottom: Peak intensity of the $^{12}$CO $J=3-2$ emission line with overlaid continuum emission (five contour levels equally placed between 3$\sigma$ and the peak flux). Note the larger spatial scale of the gas versus dust images.}
\label{fig:dust_gas}
\end{figure*}

The gas in protoplanetary disks is studied through a range of molecular lines, tracing different disk regions and probing various physical and chemical processes. While it is easier to observe substructures in the dust component of the disk due to the higher achieved sensitivity, an increasing number of substructures have also been observed in the gas (e.g. \citealp{Bruderer2014,Zhang2014,Marel2016a,Dong2017b,Law2021,Woelfer2023,Teague_exoALMA}). Similar to the dust, this includes spirals, rings, and gaps or cavities. Narrow-band line emission brings the advantage of providing information about the gas dynamics. Studying the velocity field of the rotating gas and identifying deviations from purely Keplerian rotation yields information about the underlying processes which are shaping the rotation pattern (for a review see \citealp{Pinte2022}). The different disk processes leave individual fingerprints in the kinematics but it remains difficult to disentangle them. Yet, observations of localized (e.g \citealp{Pinte2018,Pinte2019,Pinte2020,Teague2018a,Teague2019a,Casassus2019,Casassus2022,Rosotti2021,Yu2021,Alarcon2022,Izquierdo2022}) and more extended (e.g non-Keplerian spirals; \citealp{Teague2019Spiral,Teague2021,Teague2022,Rosotti2020a,Casassus2021,Garg2021,Woelfer2021,Stadler2023}) kinematical deviations from recent years are generally consistent with the presence of massive embedded planets.

Identifying vortex signatures in the gas dynamics is crucial to not only confirm their presence but also to study their dust trapping properties and understand the formation of the dust concentrations in disks. As shown by \cite{Robert2020}, vortices produce signatures potentially discernible in the line-of-sight velocity through observations with current facilities. This includes the anticyclonic motions around the density maximum of a vortex as well as vortex-driven spirals, with the latter being comparable to those excited by a planet with a mass of a few to a few tens of Earth masses \citep{Huang2019}. \cite{Yen2020} indeed report kinematical deviations in the HD\,142527 disk, which are consistent with a local pressure bump and co-located with the large horseshoe structure seen in the dust disk, yet their observations do not allow to distinguish if the pressure bump is formed by a vortex or massive planet. \cite{Boehler2021} compare CO emission lines of the same disk with a vortex model, and find kinematic signatures consistent with the presence of a large vortex, however, due to limitations in the spatial resolution of the data, artificial velocity deviations generated by beam smearing, cannot be ruled out.

In this work, we make use of the unprecedented sensitivity at high resolution of the exoALMA data \citep{Teague_exoALMA} to study the kinematical features seen around the azimuthal asymmetries of four disks in the sample and search for a kinematical vortex counterpart. In that regard, we compare the observations with both an analytical model of vortices and hydrodynamical simulations of a vortex triggered at the edge of a dead zone and gap created by a massive planet, respectively. The paper is structured as follows: in \autoref{sec:observations} we describe the source sample and present the observations. Our models are outlined in 
\renewcommand*{\sectionautorefname}{Sects.} 
\autoref{sec:methodsAna} and \ref{sec:methodsSim} and compared to the data 
\renewcommand*{\sectionautorefname}{Sect.} 
in \autoref{sec:results}. Our results are summarized in \autoref{sec:summary}.
%
%
%
%
%
%
\begin{table}[h!]
\centering
\caption{Inclination and position angle of the sources, obtained by modelling the $^{12}$CO emission line with \textsc{discminer} \citep{Izquierdo_exoALMA}. The position angle is measured counterclockwise from the north to the redshifted axis of the disk.}
\label{tab:dataInfo}
\begin{tabular}{lcccc}
\toprule
\toprule
Source & HD\,135344B & HD\,143006 & HD\,34282 & MWC\,768 \\
\midrule 
$i$ ($\degree$) & 16 & 17 & 58 & 19 \\
PA ($\degree$) & 243 & 168 & 117 & 240 \\
\bottomrule
\end{tabular}
\end{table} 
\section{Observations}\label{sec:observations}
\begin{figure*}
\centering
\includegraphics[width=1.0\textwidth]{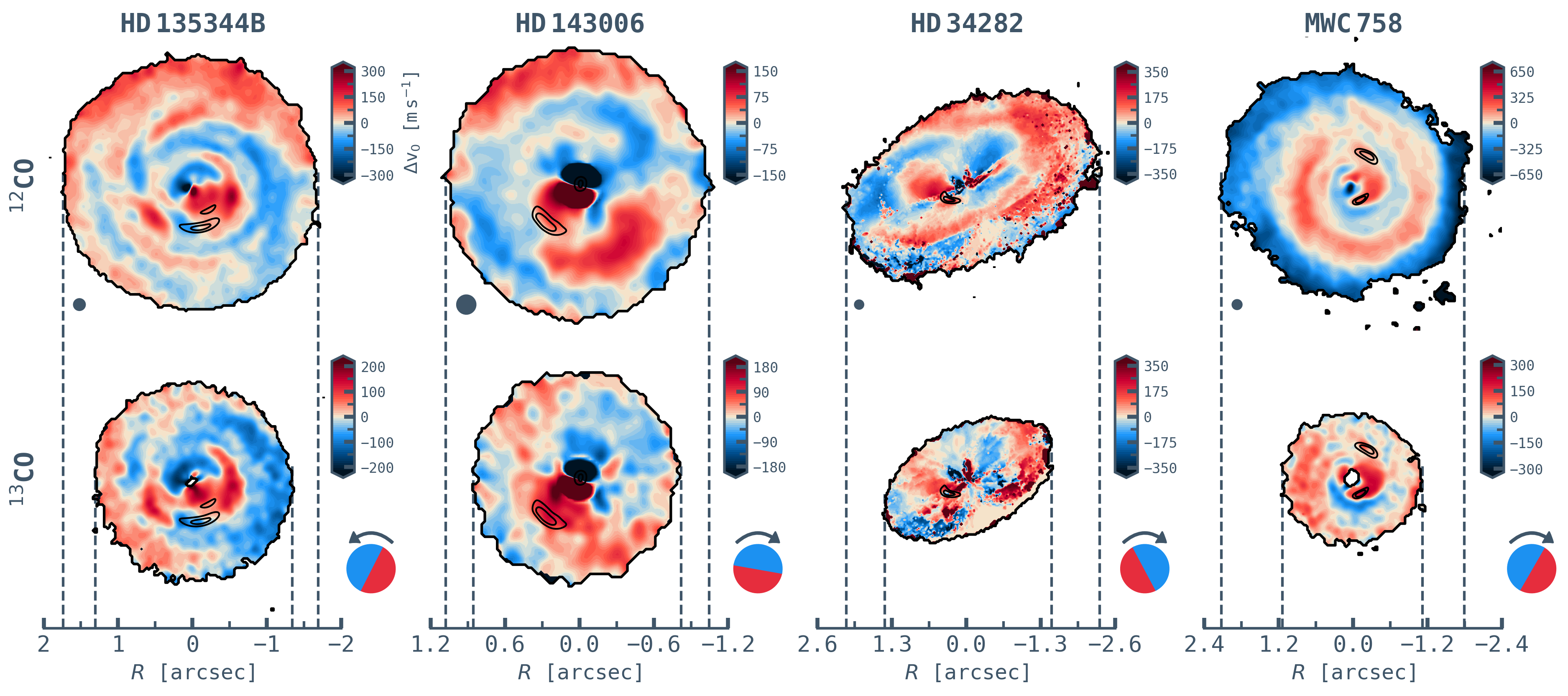}
\caption{Velocity residuals $\Delta v_0$ after subtraction of the Keplerian model from the data, shown for $^{12}$CO and $^{13}$CO $J=3-2$ and all four targets. For HD\,34282, only emission from the top side of the disk was considered when making the $v_0$ map, following the procedure described in \cite{Izquierdo_exoALMA}. The beam is shown in the bottom left corner of the top row. The red-blue circle and arrow indicate the underlying rotation pattern of each disk. To highlight the location of the crescents, two continuum contours around the dust asymmetries are overlaid on the images.}
\label{fig:VelRes}
\end{figure*}
\noindent Our source sample consists of four sources -- namely HD\,135344B, HD\,143006, HD\,34282, and MWC\,758 -- which were observed as part of the exoALMA large program \citep{Teague_exoALMA}. They are among the six most asymmetric disks in continuum \citep{Curone_exoALMA} in the exoALMA sample and all show evidence of azimuthal dust traps in their millimetre (mm) continuum (compare top row of \autoref{fig:dust_gas}), possibly caused by the presence of a vortex. Three of the sources -- HD\,135344B, HD\,34282, and MWC\,758 -- are further marked by spiral structures in near-infrared (NIR) scattered light (compare middle row of \autoref{fig:dust_gas}, \citealp{Stolker2017,Boer2021,Ren2023}). A summary of each source is provided \autoref{sec:target_summary}. The inclination and position angle of the targets, obtained with \textsc{discminer} \citep{Izquierdo_exoALMA}, are summarized in \autoref{tab:dataInfo}.
%
%
%
%
%

To study the gas features seen around the dust azimuthal asymmetries in our sample, we make use of the fiducial exoALMA images \citep{Teague_exoALMA} of the $^{12}$CO and $^{13}$CO $J=3-2$ emission lines (0\farcsec15;  100$\,$m$\,$s$^{-1}$, continuum-subtracted). The CS $J = 7-6$ images are excluded from this work due to their much lower SNR at the same resolution. For the full details of the imaging techniques, we refer the reader to \cite{Loomis_exoALMA}. For both lines, the molecular emission channels were modeled with the \textsc{discminer} package \citep{Izquierdo2021}, which returns a smooth Keplerian model of the line intensity profile. To identify perturbations in the gas dynamics, it is then necessary to extract observables - such as the centroid velocity, line peak intensity, and line width - in a post-processing step and to compute deviations between the data and model (residuals). Here, this is done by fitting both the data and model line profiles with a Gaussian function, except for the HD\,34282 disk for which a double-peaked bell profile is adopted. While for low-inclination disks, a single-peak profile is sufficient to represent the morphology of the profile, higher-inclination disks such as HD\,34282 are better reproduced with a double-peak profile since emission from the back side of the disk is visible. This approach then allows to separate the contribution from the front and back sides of the disk. Details of the modeling procedures and choice of extraction methods can be found in \cite{Izquierdo_exoALMA}.

In the bottom row of \autoref{fig:dust_gas}, a gallery of the $^{12}$CO peak intensity is shown for the four targets studied here. Emission below 5$\sigma$ is masked. The continuum emission (top row) is overlaid as contours, placed equally between 3$\sigma$ and the continuum peak, highlighting the location of the potential vortices at relatively small radii compared to the large extents of the gas disk. The latter exceeds the dust disk by a factor of roughly 2-3, which is expected from processes such as radial drift or the difference between dust and gas opacities \citep{Facchini2017,Trapman2019}. Given that $^{12}$CO emission is optically thick, the peak intensity traces the temperature of the gas, with the observed temperatures of up to $\sim 100$\,K being as expected in the upper disk layers (e.g. \citealp{Bruderer2013,Bruderer2014, Leemker2022}). A thorough analysis of the 2-dimensional temperature structure of the disks can be found in \cite{Galloway_exoALMA}.
\subsection{Velocity residuals}
\noindent In \autoref{fig:VelRes}, we present the velocity centroid residuals of our targets after subtraction of the Keplerian \textsc{discminer} model from the $^{12}$CO (top row) and $^{13}$CO (bottom row) data (Fukagawa et al. in prep.), respectively. 
For the HD\,34282 disk, the contribution from the back side has been subtracted in the computation of the residuals, as discussed in \cite{Izquierdo_exoALMA}. The other three disks are close to face on and do not require a removal of the lower surface.
\begin{figure*}
\centering
\includegraphics[width=1.0\textwidth]{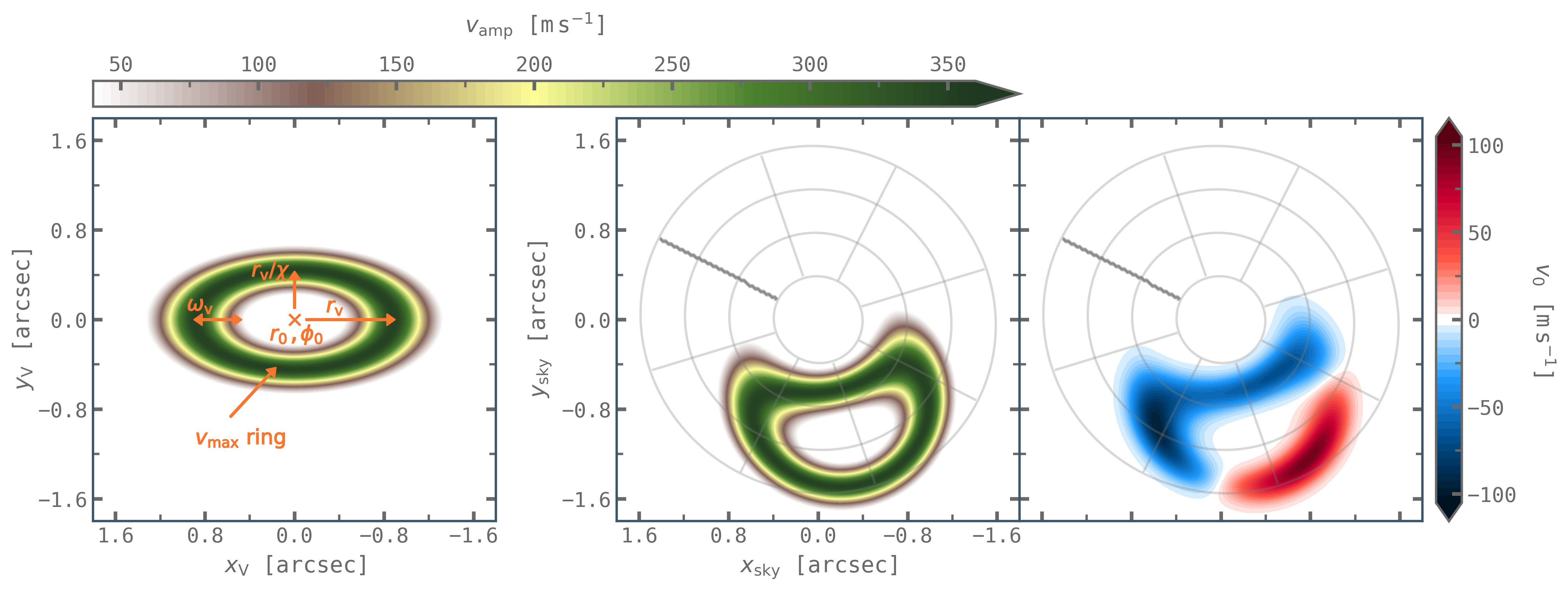}
\caption{Analytical model of a vortex, described as elliptic streamlines of constant velocity with a Gaussian velocity profile. The first two panels show the absolute velocity of the vortex in both the vortex frame and sky frame. The last panel depicts the velocity projected along the line-of-sight, showing a complex morphology purely resulting from the projection. In the last two panels, the underlying disk geometry is overlaid as grey contours, corresponding to a position angle of $243\,\degree$ and inclination of $16\,\degree$ (same as for HD\,135344B).}
\label{fig:vortexAnalytical}
\end{figure*}

The HD\,135344B kinematics are marked by prominent asymmetric arc- and spiral-like structures, seen in both emission lines, as well as features resembling a Doppler flip. Similarly, MWC\,758 shows a large spiral arm, covering more than one full azimuth. For both disks, similar features are seen in the peak intensities and line widths (compare \autoref{fig:polarSpirals}). Altogether, this suggest strong perturbations in these disks. Interestingly, they are also the two disks in our sample that exhibit extended spirals in the near-infrared scattered light (see \autoref{sec:spirals}). The patterns seen for the HD\,143006 disk are less clear or consistent, but a prominent red-shifted arc can be distinguished in the kinematics traced by both lines (even though at an offset), which may connect to the extended red-shifted arc in the outer disk. The residuals of the HD\,34282 disk are particularly difficult to interpret due to the high-inclination nature of the source. Even though the \textsc{discminer} model and subsequent fit with a double-peaked bell profile account for both the upper and lower disk surface, contributions from the lower side seem to still be visible in the upper surface residuals and the vertical structure off the disk is seen as a butterfly pattern in the inner disk regions. 
\section{Methods I: Analytical Vortex Model}\label{sec:methodsAna}
\noindent In the following sections, we model the kinematical signature of a vortex to compare with the deviations seen in the residuals of our sample. For that purpose, we first construct a simple analytical model similar to \cite{Boehler2021}, which serves as a guideline for understanding the general kinematical patterns triggered by a vortex, and helps to gain intuition for the more complex residuals. Subsequently, we perform 2-dimensional hydrodynamical simulations (see \autoref{sec:methodsSim}). 
Doing so, we are mostly interested in comparing the kinematical patterns of a vortex to the data rather than a full representation of the vortex properties, which remains challenging \citep{Huang2019,Robert2020}. 
\begin{figure*}
\centering
\includegraphics[width=1.0\textwidth]{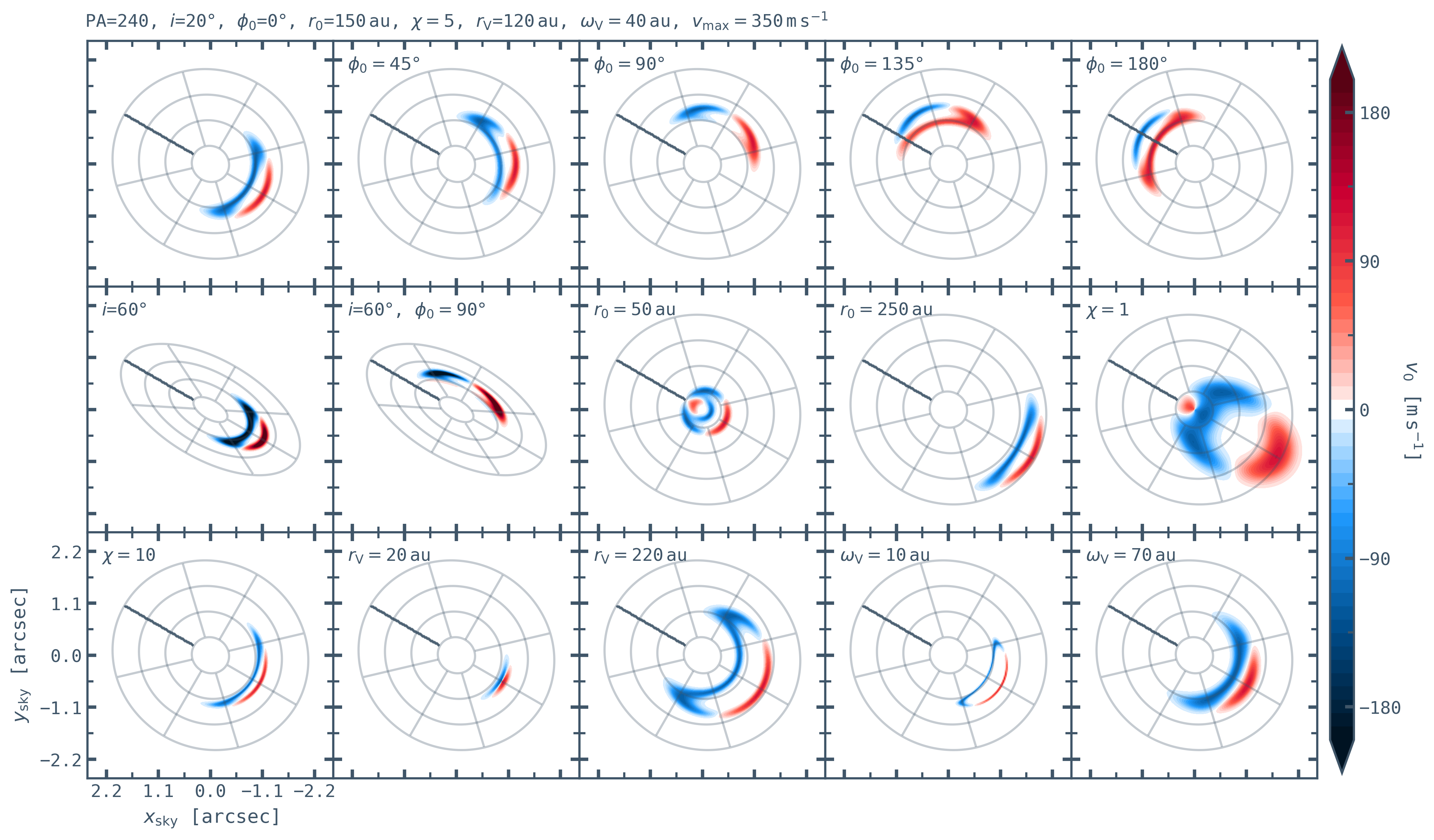}
\caption{Line-of-sight velocity of the analytical vortex model, shown for different disk geometries and vortex parameters. The parameters given on top of the figure correspond to the morphology shown in the first panel. In the top of each individual panel, the parameter changed compared to this base vortex model are given. The underlying disk geometry is overlaid as grey contours.}
\label{fig:vortexAnalytical2}
\end{figure*}
\subsection{Model description}\label{sec:AnalyticalSetup}

\noindent We follow \cite{Boehler2021} in setting up the vortex flow as streamlines of constant velocity around a central position (r$_0$, $\phi_0$) in polar disk coordinates, 
with the rotation velocity of the vortex varying as a Gaussian in the radial direction. In contrast to \cite{Boehler2021}, the streamlines are elliptical in the vortex frame in Cartesian coordinates, analogous to a shearing-box. Our parameterization is depicted in the first panel of \autoref{fig:vortexAnalytical}. Here, the vortex is shown in the Cartesian vortex frame (subscript v) and is thus centered around the origin of this frame. The velocity in the vortex increases to a maximum velocity $v_{\mathrm{max}}$ at the radius $r_{\mathrm{v}}$. The vortex is further described by an aspect ratio $\chi$, the ratio of the major axis in the azimuthal direction to the minor axis in the radial direction.
The width of the Gaussian profile is given by $\omega_{\mathrm{V}}$. 

The Cartesian vortex coordinates $(x_{\mathrm{V}},y_{\mathrm{V}})$ can be related to the polar disk coordinates $(r,\phi)$ through
\begin{align}
& x_{\mathrm{V}} = r (\phi - \phi_0)\,\,\&\\
& y_{\mathrm{V}} = r-r_0,
\end{align}
where $(r_0,\phi_0)$ describes the center of the vortex. These can further be transformed into polar vortex coordinates $(r',\phi')$ via 
\begin{align}
& r' = \sqrt{x_{\mathrm{V}}^2 + (\chi \cdot y_{\mathrm{V}})^2}\,\,\& \\
& \phi' = \tan^{-1}\left(\frac{x_{\mathrm{V}}}{\chi \cdot y_{\mathrm{V}}}\right). 
\end{align}
The absolute rotation velocity of the Vortex is then given as
\begin{equation}
|v(x_{\mathrm{V}},y_{\mathrm{V}})| = v_\mathrm{max} \cdot \exp \left[\left(\frac{r'-r_{\mathrm{V}}}{\omega_{\mathrm{V}}}\right)^2\right].    
\end{equation}

In \autoref{fig:vortexAnalytical}, the absolute orbital velocity is shown in the vortex frame (first panel) alongside the sky frame (second panel), using an inclination of $16\,\degree$ and position angle of $243\,\degree$ for the projection (same as for HD\,135344B). The velocity projected along the line-of-sight is then given as
\begin{equation}
v_0 = |v(x,y)| \cdot \cos{(\phi'+\phi)} \cdot \sin{i},    
\end{equation}
where $i$ represents the disk inclination. The projected velocity of the vortex is presented in the last panel of \autoref{fig:vortexAnalytical} in the sky frame, showing a clear red-blue pattern with a complex morphology, which purely results from projection effects. 
\subsection{Parameter exploration}
\noindent In \autoref{fig:vortexAnalytical2}, we present the line-of-sight velocity for different disk geometries as well as parameters of the analytical vortex model. In this context, the parameters given on top of the figure correspond to the base model shown in the first panel, while the individual parameters changed compared to this model are given in each panel. This exercise highlights that while generally, a similar red-blue pattern is visible in all cases, the exact morphology and strength of the pattern varies substantially for different combinations of parameters. We further note that the vortex edges have strong radial components, while along the `arcs', the deviations are dominated by azimuthal motions.

\noindent We attempted to constrain the basic properties of the dust crescents seen in our targets by implementing the simple vortex model in the \texttt{eddy} code \citep{eddy}, which allows to fit a Keplerian profile to a rotation map. In this context, we first conducted a run to find a good representation of the main rotation pattern, and in subsequent runs fixed those values to only fit for the vortex parameters described in \autoref{sec:AnalyticalSetup}, fitting different disk regions. However, none of these runs converged and resulted in very degenerate distributions. Given the complex kinematical patterns seen in our targets (see \autoref{fig:VelRes}), we conclude that a simple model is not sufficient to constrain the properties of the vortex. We are further limited by the low inclination of three of the four sources, making it hard to detect azimuthal motions projected into the line-of-sight. This is different for vertical motions. The importance of vertical motions is unclear at this point (e.g., \citealp{Meheut2010,Meheut2012a,Lin2012,Richard2013}), but they may dominate the vortex signature in some cases, which cannot be captured by our simple 2D model. Moreover, other disk processes such as planet-disk interactions can create dynamical patterns which overlay the vortex signatures and make it challenging to isolate them (see also \citealp{Barraza2024}). Nevertheless, the analytical model is useful to gain intuition for the general patterns created by a vortex and can help to interpret both the simulations and observations (see \autoref{sec:methodsSim}).
\begin{figure*}
\centering
\includegraphics[width=1.0\textwidth]{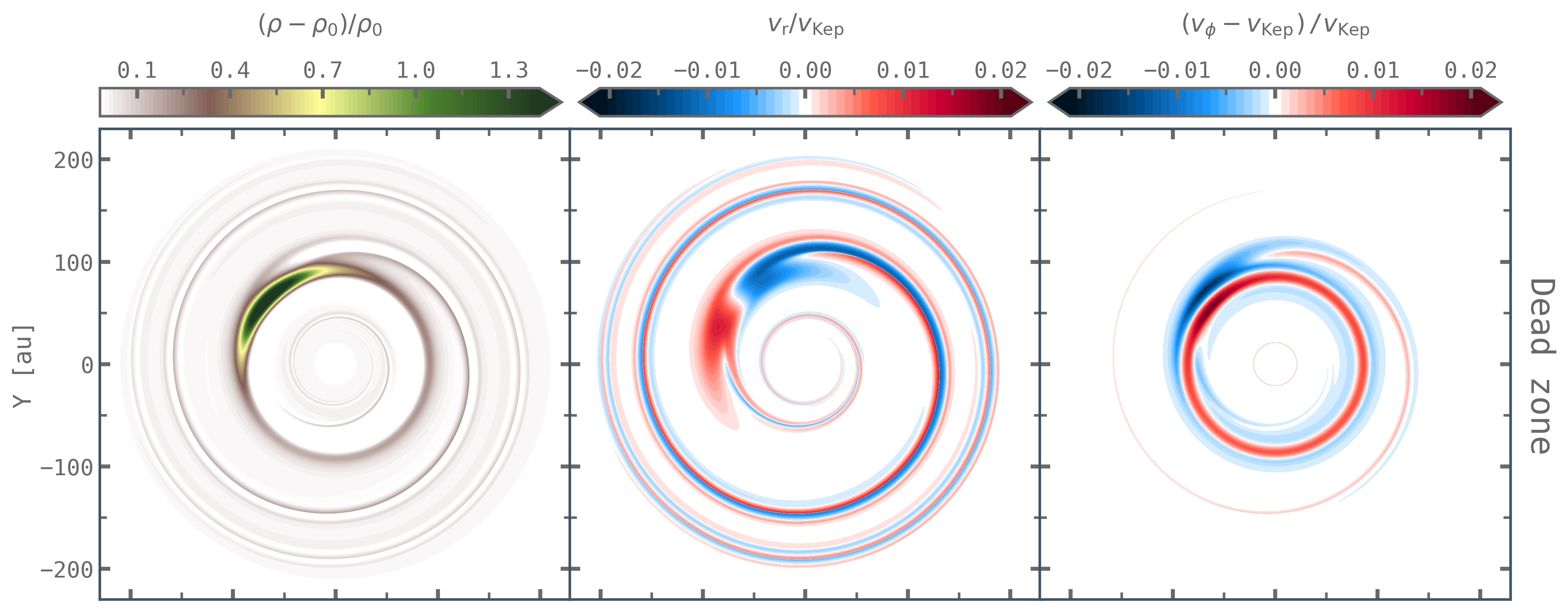}
\includegraphics[width=1.0\textwidth]{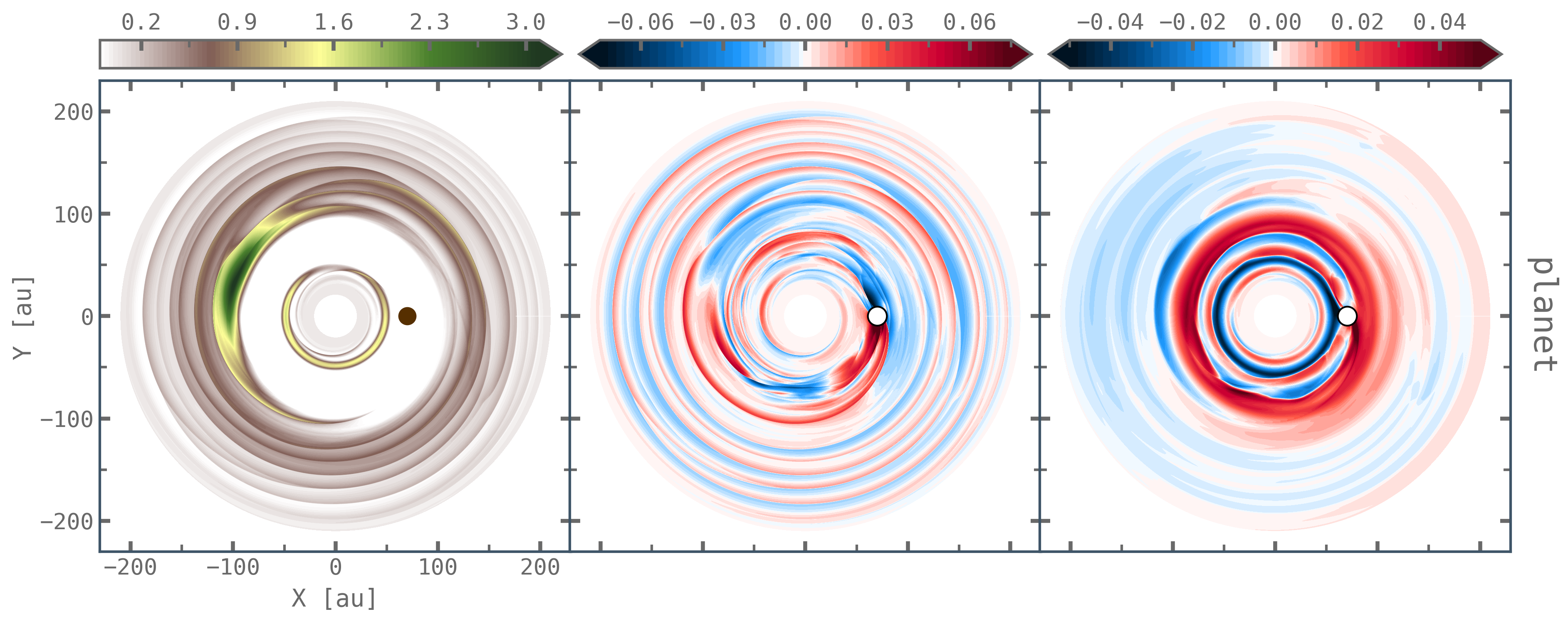}
\caption{\textsc{PLUTO} simulations of a vortex triggered by the RWI at the edge of a dead zone (top row) and a Jupiter-mass planet (bottom row), using a scaling radius of 70\,au to convert from code units to physical units. Shown are the gas density (left panel), radial velocity (middle panel), and azimuthal velocity (right panel). The density field and sub-Keplerian rotation velocity of a smooth disk have been removed (in case of $v_{\mathrm{r}}$ this is zero) and the quantities are normalized. The location of the planet is marked with a circle in the bottom row panels.}
\label{fig:vortexSimulation}
\end{figure*}
\section{Methods II: Simulations}\label{sec:methodsSim}
\noindent We explore the kinematic structure from a vortex (and vortex+planet) by conducting 2D numerical simulations with the grid-based code \textsc{PLUTO} \citep{Mignone2007}, version 4.4. In this context, we do not aim for a comprehensive parameter study but intend to expand beyond the relatively simple analytical model to try to capture some of the complexity seen in the data. We follow the disk evolution, solving the Navier-Stokes equations of fluid dynamics (for details see Section 2 in \citealt{Barraza2024}) for two scenarios: a vortex triggered by the Rossby-wave instability (RWI) at the inner edge of a MRI dead-zone, and RWI triggered at the edge of a gap opened by a Jupiter mass planet. Even though the inner edge of the dead zone -- in contrast to the outer edge -- is unlikely to be relevant on exoALMA scales, we set up the simulation in this way to trigger a vortex at a particular location. The mechanism generating the vortex is the same for both the inner and outer edge, but in the latter case results in a less clean vortex and uncertain location due to radial migration. Here we are mainly interested in the morphology of the vortex and creating a vortex in the same location for both the dead zone and embedded planet simulation makes it easier to compare the two cases. In both cases, a vortex spontaneously forms within a few hundred orbits, computed at 1 code unit length in the dead zone scenario and corresponding to planetary orbits in the planetary case.

We use a \textsc{PLUTO} numerical setup with LINEAR reconstruction method, 2nd order Runge-Kutta time-stepping, the Harten-Lax-Van Leer Riemann (HLLC) solver, and we include the disk viscosity as a diffusion term integrated with Super-Time-Stepping (STS) technique. We set the Courant number (CFL) to 0.4.
\subsection{Disk Setup}
\noindent We consider a two-dimensional disk in polar coordinates, including the evolution of the gas only. The disk initial conditions are based on the numerical setup presented in \cite{Miranda2017}. Our simulations follow a locally isothermal equation of state $P=c_{\mathrm{s}}^2 \Sigma$, where $P$ is the vertically-integrated gas pressure, and the disk gas surface density $\Sigma_{\mathrm{g}}$ follows:
\begin{equation}
    \Sigma_{\mathrm{g}} = \Sigma_0 \left(\frac{r}{r_0}\right)^{-1},
\end{equation}
with $r_0$ the code unit of length, and $\Sigma_0$ the gas surface density at $r=r_0$. The disk sound speed $c_{\mathrm{s}}$ has a radial dependency $c_{\mathrm{s}} = c_{\mathrm{s},0}(r/r_0)^{-0.25}$. The value of $c_{\mathrm{s},0}$ is set so that $H(r_0)/r_0=0.05$, with $H$ the disk pressure scale height, $H=c_s/\Omega_K$, where $\Omega_K = \sqrt{G M_{\star}/r^3}$ is the Keplerian orbital frequency. Our disk model results in a flared disk, with $H\propto r^{1.25}$. 

The simulation domain covers 0.3-3.0 code units of length in the radial direction, and $2 \pi$ rad in azimuth. The simulations were run with a grid of $1024$ and $2048$ cells in $r$ and $\phi$, respectively, with a mesh uniformly spaced in both the radial and azimuthal directions. The grid uniform radial spacing allows a better resolved dead-zone edge transition and outer planetary gap edge relative to a logarithmic mesh.
The effects of disk self-gravity are not taken into account in our models.
\subsubsection{Dead Zone Edge}
\noindent To mimic a MRI-deadzone transition, we follow \cite{Miranda2017}, however, for the inner edge of a dead-zone instead of the outer one, implementing a transition between two levels of viscosity described by a radially-dependent $\alpha$-viscosity model, $\nu=\alpha c_{\mathrm{s}} H$ \citep{Shakura1973}, with $\alpha$ following
\begin{equation}
    \alpha(r) = \alpha_0 \left( 1 - \frac{1}{2} \left( 1 - \frac{\alpha_{\mathrm{DZ}}}{\alpha_0} \right) \left[ 1 + \tanh{\left( \frac{r-r_{\mathrm{DZ}}}{\Delta_{\mathrm{DZ}}} \right)} \right]  \right),
\end{equation}
with $\alpha_{\mathrm{DZ}}=10^{-5}$ and $\alpha_{0}=10^{-3}$ being the $\alpha$-viscosity parameters in the MRI-dead and MRI-active regions, respectively. The location of the transition between the active and dead zones is put as 1.2 code units of length ($r_{\mathrm{DZ}}=1.2 r_0$), and $\Delta_{\mathrm{DZ}}$ sets the width of the transition. We set $\Delta_{\mathrm{DZ}}=H(r_{\mathrm{DZ}})$, corresponding to the standard run from \cite{Miranda2017} (see their Table 1).
\begin{figure*}
\centering
\includegraphics[width=1.0\textwidth]{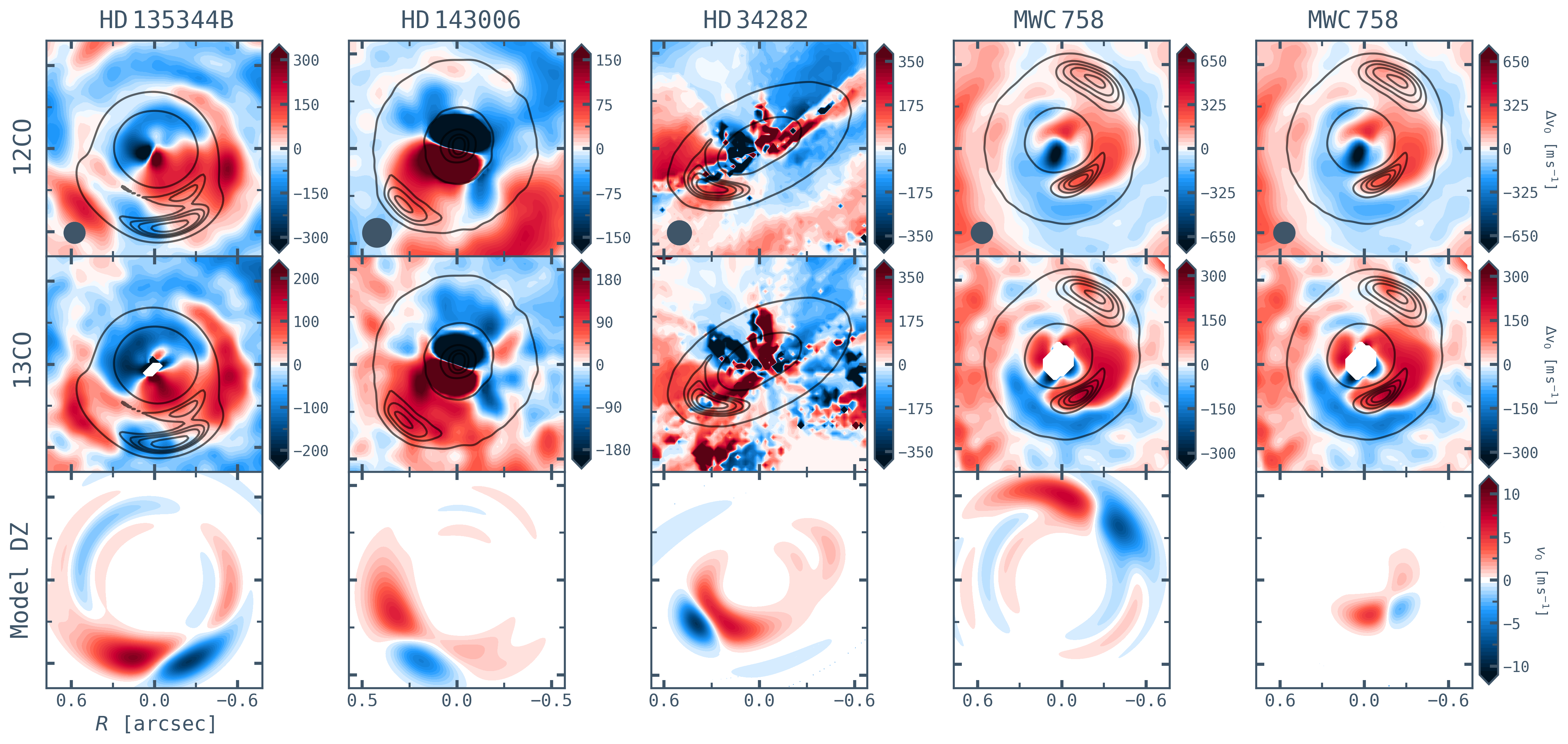}
\caption{Velocity residuals of the data (top two rows) and dead zone vortex model (bottom row). The model was post-processed with RADMC-3D and convolved with the beam of the observation. For each disk, the model has been adapted such that that the geometry and vortex location match the data. For MWC\,758, two representations are included to account for both vortex features. In the first two rows, the continuum observations are overlaid, highlighting the dust asymmetries. 
The beam of the observation is shown in the lower left corner of each panel in the first row.}
\label{fig:vortexCompare}
\end{figure*}
\subsection{Planet}
\noindent In a second setup, we include a $1\,M_{\mathrm{J}}$ planet and the star as the gravitational potential of points of mass, and the indirect acceleration term
\begin{equation}
    \Phi = \frac{G M_{\star}}{r} - \frac{G m_{\mathrm{p}}}{\sqrt{|\vec{r}-\vec{r_{\mathrm{p}}}|^2+s^2}} + \frac{G m_{\mathrm{p}}}{r_{\mathrm{p}}^2}r \cos{\phi},
\end{equation}
where $\vec{r_{\mathrm{p}}}$ is the distance from the star to the planet, and $s$ a potential smoothing length set to $60\%$ of the local disk pressure scale height (see e.g, \citealt{Masset2002, Muller2012, Weber2019}). We fix the planet orbital distance to 1 code unit of length (or $r_{\mathrm{p}}=r_0$). In addition, we include a tapering such that the planet mass smoothly reaches its final value in $30$ planetary orbits. Finally, for the planet simulation, we use a constant $\alpha$ viscosity value of $10^{-5}$.
\\
\\
\noindent In \autoref{fig:vortexSimulation}, we show the density (left panel), radial velocity (middle panel), and azimuthal velocity relative to a sub-Keplerian background (right panel) resulting from the simulations of a vortex triggered by the RWI at the edge of a dead zone (top row) and gap created by a massive planet (bottom row), respectively. We have used a scaling radius of 70\,au to convert from code units to physical units (thus $r_{\mathrm{DZ}} = 84\,\mathrm{au}$ and $r_{\mathrm{p}} = 70\,\mathrm{au}$). To highlight the patterns created by the vortex, we have subtracted the density field and sub-Keplerian rotation velocity of a smooth disk from the perturbed density and velocity fields. Since in this work we are primarily interested in the patterns created by a vortex and not the exact magnitude of the perturbation, we have further normalized the quantities by the local sub-Keplerian velocity. The simulation snapshots depicted correspond to 400 orbits measured at 1 code unit of length (dead zone) and 580 planetary orbits, respectively.  
\section{Results}\label{sec:results}
\noindent Similar to previous works (e.g \citealp{Huang2019,Robert2020}), it is apparent that the vortex creates a prominent azimuthally asymmetric feature in the gas dynamics, seen as a clear red-blue dipole pattern, as well as spiral structures
\renewcommand*{\figureautorefname}{Figs.}(\autoref{fig:vortexAnalytical2} and \ref{fig:vortexSimulation}).
\renewcommand*{\figureautorefname}{Fig.}
The vortex motion looks similar in both the simulations and analytical model, but the spiral features cannot be captured by the latter. If a planet is included in the disk, it creates its own Doppler-flip signature and much stronger spiral features, which overlay on the pattern associated with the vortex, thus making it much harder to distinguish.

To compare the vortex signatures predicted by the simulations with the features seen in the velocity residuals, we post-processed the dead zone simulation with RADMC-3D \citep{Dullemond2012} and computed mock images of a flat disk model. In this context, we took the geometry of the sources into account and then matched the radial and azimuthal location of the vortex with the dust asymmetries seen in the data. To directly obtain the expected residuals from the simulated vortex dynamics, we subtracted the sub-Keplerian background from the azimuthal velocity used as an input in the radiative transfer model. We then created synthetic $^{12}$CO channel maps via ray-tracing with the same spectral resolution as the data. As a final step, we convolved the synthetic data cubes with a 0\farcsec15 Gaussian beam, and computed the line-of-sight velocity maps using \textsc{bettermoments} \citep{bettermoments}. 

The result of this test is shown in \autoref{fig:vortexCompare}, where the first two rows represent the $^{12}$CO and $^{13}$CO residuals, and the bottom row the dead zone vortex simulation. The continuum emission is overlaid on the data to highlight the location of the dust crescents. For MWC\,758, two representations are included since this disk is marked by several azimuthal asymmetries. The unconvolved model images are shown in \autoref{sec:model_unconv}.

Comparing the data and model, it becomes apparent that most disks do not show a clear vortex kinematic signature around the dust crescents as predicted by the simulation. In HD\,135344B, a similar red-blue pattern is traced by both emission lines, yet it is shifted with respect to the dust asymmetry and, overlaid by the strong spiral features, and nearby the disk minor axis, where a change of sign due to coherent azimuthal variations is expected (Fig.5 in \citealp{Teague2019Spiral}). Even though the vortex also triggers a similar spiral pattern, it is generally not expected to be strong enough to distinguish it in the kinematics (see also \citealp{Huang2019}). Close to the inner crescent in MWC\,758, a red-blue flip is present which resembles the vortex model, but it is part of the red and blue spiral arms and less distinct than the red-blue pair spanning the dust asymmetry in HD\,135344B. Among our sample, this makes HD\,135344B the most convincing example of a vortex and the best target to pursue with even deeper observations at very high angular resolution. Connected to that, when comparing the size of the dust asymmetries with the size of the beam, it becomes clear that they are very similar, thus the vortex signatures may not be resolved and therefore do not clearly show up in the kinematics. 

Aside from planet-disc interactions, other processes such as (magneto-)hydrodynamical instabilities and the gravitational instability \citep{Barraza2024,Barraza_exoALMA,Hall2020}, stellar interactions including binaries \citep{Calcino2020,Norfolk2022} and fly-bys \citep{Cuello2020}, or magnetic winds \citep{Galloway2023} can contribute to the kinematics and may obscure the signatures from a vortex, resulting in complex patterns which are hard to disentangle. \cite{Ragusa2024} make predictions for the kinematics expected from an eccentric circumbinary disk, finding that they result in a single lobed pattern at the cavity edge. Compared to our data, we cannot distinguish such a pattern, though this may result from resolution effects.   
%
%
\section{Summary \& Conclusions}\label{sec:summary}
\noindent In this work, we have studied four disks within the exoALMA sample which show evidence of dust trapping in a vortex, to search for a kinematic counterpart of the latter. Our main results are summarized as follows.
\begin{itemize}
\item The kinematics of our targets, which all show azimuthal dust asymmetries, are marked by complex patterns, including spirals and Doppler flip features (\autoref{fig:VelRes}). Two of the sources, HD\,135344B and MWC\,758, show prominent spirals in both the NIR scattered light and all three residuals, with the substructures partially coinciding (\autoref{sec:spirals}). 
\item An analytical model of the velocity perturbation from a vortex reveals a distinct red-blue pattern but shows different morphologies depending on the underlying disk geometry and exact vortex parameters \renewcommand*{\figureautorefname}{Figs.}
(\autoref{fig:vortexAnalytical},\ref{fig:vortexAnalytical2}). \renewcommand*{\figureautorefname}{Fig.}
This simple model is, however, insufficient to explain the complexity of structure seen in the data.  
\item Hydrodynamical simulations of a vortex and vortex+planet also reveal a clear red-blue kinematical pattern in addition to spiral features. In the planetary scenario, the signatures of the vortex are mixed with the strong perturbations from the planet, making it much more difficult to distinguish them (\autoref{fig:vortexSimulation}). Other disk processes perturbing the gas dynamics are expected to further complicate this picture. 
\item Comparing the kinematical patterns of the data and model shows that the features seen around the dust crescents cannot unambiguously be linked to a vortex (\autoref{fig:vortexCompare}). Aside from the complications arising due to different physical processes impacting the kinematics, our results may also limited be by the spatial resolution in comparison to the size of the dust asymmetries. While fully resolving the vortex is not necessarily needed to trace the red-blue flip, it is essential if we are to parameterize it and trace the pressure-profile across the vortex. As shown by \cite{Stadler_exoALMA}, the pressure morphology of dust rings can only be inferred if they are spatially resolved and we are faced with same challenge in the context of vortices.     
\end{itemize}
While it is necessary to aim for higher spatial resolution to unambiguously detect the kinematic signature of a vortex, it can also be useful to study emission lines which trace closer to the midplane, where vortices reside, to be more sensitive to the vortex dynamics (compare \citealp{Pinte2019}, showing how the strength of a planet's perturbation diminishes with height). This is the case for CS $J=7-6$ (compare \citealp{Galloway_exoALMA}), which was observed within the exoALMA program (Cataldi et al. in prep.). Both the SNR at high spatial resolution and likely the resolution itself are, however, insufficient to draw conclusions about the kinematical pattern around the dust asymmetries, but could be studied in a future work. 
Getting as close as possible to the midplane may increase our chance of detecting the vortex in the gas dynamics significantly. 

The vortex predictions presented in this work rely on 2D models. However, to reveal the full dynamics of a vortex and in particular understand the role of vertical motions, global 3D simulations are essential. These simulations are challenging and the formation and stability of 3-dimensional vortices remains an active matter of debate. Earlier works have addressed the question of 2D versus 3D vortices triggered by the RWI, but have resulted in different conclusions: While the general vortex patters are consistent between 2D and 3D simulations, some studies suggest vertical motions to be negligible \citep{Lin2012,Richard2013}, others show that they can be more significant in certain cases \citep{Meheut2010,Meheut2012a}. More detailed theoretical studies are needed to understand how the strength of features varies with height and if they may be reliably detectable in tracers such as $^{12}$CO and $^{13}$CO.    

Within the exoALMA sample, the most promising source for follow-up observations to detect the kinematic signature of an anti-cyclonic vortex is HD\,135344B, as it is marked by a relatively large dust crescent and the kinematics show a pattern which resembles the signatures of a vortex. In terms of the expected strength of the kinematic signature from a vortex, HD\,34282 may be another useful target to follow-up with higher spatial resolution and less abundant tracers at higher sensitivity due to its large inclination. The dust asymmetry in this source is, however, relatively small and located closer to the star, making such observations challenging. 


\clearpage
\section*{Acknowledgments}
\noindent We thank the anonymous referee for the comments, which greatly improved the quality of this paper.
%
This paper makes use of the following ALMA data: ADS/JAO.ALMA\#2021.1.01123.L. ALMA is a partnership of ESO (representing its member states), NSF (USA) and NINS (Japan), together with NRC (Canada), MOST and ASIAA (Taiwan), and KASI (Republic of Korea), in cooperation with the Republic of Chile. The Joint ALMA Observatory is operated by ESO, AUI/NRAO and NAOJ. The National Radio Astronomy Observatory is a facility of the National Science Foundation operated under cooperative agreement by Associated Universities, Inc. We thank the North American ALMA Science Center (NAASC) for their generous support including providing computing facilities and financial support for student attendance at workshops and publications. 

%
The authors acknowledge the MIT SuperCloud and Lincoln Laboratory Supercomputing Center for providing HPC resources that have contributed to the research results reported within this paper.

JB acknowledges support from NASA XRP grant No. 80NSSC23K1312. MB, DF, JS have received funding from the European Research Council (ERC) under the European Union's Horizon 2020 research and innovation programme (PROTOPLANETS, grant agreement No. 101002188). Computations by JS have been performed on the 'Mesocentre SIGAMM' machine, hosted by Observatoire de la Cote d'Azur. PC acknowledges support by the Italian Ministero dell'Istruzione, Universit\`a e Ricerca through the grant Progetti Premiali 2012 – iALMA (CUP C52I13000140001) and by the ANID BASAL project FB210003. SF is funded by the European Union (ERC, UNVEIL, 101076613), and acknowledges financial contribution from PRIN-MUR 2022YP5ACE. MF is supported by a Grant-in-Aid from the Japan Society for the Promotion of Science (KAKENHI: No. JP22H01274). CH acknowledges support from NSF AAG grant No. 2407679. JDI acknowledges support from an STFC Ernest Rutherford Fellowship (ST/W004119/1) and a University Academic Fellowship from the University of Leeds. Support for AFI was provided by NASA through the NASA Hubble Fellowship grant No. HST-HF2-51532.001-A awarded by the Space Telescope Science Institute, which is operated by the Association of Universities for Research in Astronomy, Inc., for NASA, under contract NAS5-26555. CL has received funding from the European Union's Horizon 2020 research and innovation program under the Marie Sklodowska-Curie grant agreement No. 823823 (DUSTBUSTERS) and by the UK Science and Technology research Council (STFC) via the consolidated grant ST/W000997/1. FMe acknowledges funding from the European Research Council (ERC) under the European Union's Horizon Europe research and innovation program (grant agreement No. 101053020, project Dust2Planets). CP acknowledges Australian Research Council funding via FT170100040, DP18010423, DP220103767, and DP240103290. DP acknowledges Australian Research Council funding via DP18010423, DP220103767, and DP240103290. GR acknowledges funding from the Fondazione Cariplo, grant no. 2022-1217, and the European Research Council (ERC) under the European Union’s Horizon Europe Research \& Innovation Programme under grant agreement no. 101039651 (DiscEvol). H-WY acknowledges support from National Science and Technology Council (NSTC) in Taiwan through grant NSTC 113-2112-M-001-035- and from the Academia Sinica Career Development Award (AS-CDA-111-M03). GWF acknowledges support from the European Research Council (ERC) under the European Union Horizon 2020 research and innovation program (Grant agreement no. 815559 (MHDiscs)). GWF was granted access to the HPC resources of IDRIS under the allocation A0120402231 made by GENCI. AJW has received funding from the European Union’s Horizon 2020 research and innovation programme under the Marie Skłodowska-Curie grant agreement No 101104656. TCY acknowledges support by Grant-in-Aid for JSPS Fellows JP23KJ1008. Support for BZ was provided by The Brinson Foundation. Views and opinions expressed by ERC-funded scientists are however those of the author(s) only and do not necessarily reflect those of the European Union or the European Research Council. Neither the European Union nor the granting authority can be held responsible for them. 
\clearpage
\appendix
\section{Summary of the sources}\label{sec:target_summary}
\subsection*{HD\,135344B}
\noindent HD\,135344B -- or SAO\,206462 -- is a young (8.9\,Myr; \citealp{Asensio2021}) Herbig F-type star located in the Upper Centaurus-Lupus star-forming region at a distance of $\sim 135\,\mathrm{pc}$ (Gaia DR3; \citealp {Gaia2023}). It is part of a visual binary system (separation 21\arcsec, i.e. $\sim$ 2800\,au; \citealp{Mason2001}) and surrounded by a low-inclination transition disk which has been extensively studied at different wavelengths. 

The disk was first resolved with the \textit{Hubble Space Telescope} \citep{Grady2009}, with subsequent scattered light observations revealing substructures such as an inner dust cavity ($\sim$ 28\,au) and two prominent spiral arms on a scale of $\sim$ 0.2-0.6\arcsec \citep{Muto2012,Garufi2013,Stolker2016,Stolker2017}, one being marked by a brightness asymmetry and pitch angle change, possibly caused by the spiral moving through a high-density vortex region \citep{Bae2016}. Moreover, the scattered light images exhibit shadow features which may be related to misaligned inner disk regions \citep{Stolker2016,Stolker2017}.  

A large dust cavity ($\sim$ 50\,au), originally suggested through SED modelling by \citealp{Brown2007}, was later confirmed in the mm-continuum with the SMA  \citep{Brown2009,Andrews2011} and ALMA \citep{Perez2014,Marel2016,Cazzoletti2018}. Additionally, these data revealed an asymmetric crescent located further out ($\sim$ 80\,au), which is marked by an azimuthal peak shift with wavelength and matches the predictions of a dust-trapping vortex \citep{Marel2016,Cazzoletti2018}. The two dust rings are connected by a filament, which co-locates with the deviation in the spiral arm and may be tracing a planetary wake crossing the dust gap \citep{Casassus2021}. 

Observations of CO gas in the disk revealed an inner gas cavity peaking inside the dust cavity ($\sim$ 30\,au) and a spiral feature in the kinematics \citep{Marel2016,Casassus2021}. Similar to the NIR data, shadows are seen in archival C$^{18}$O and exoALMA CS data (Cataldi et al. in prep.). Several works have shown that the different observations may be explained by one or more massive embedded planets (e.g. \citealp{Muto2012,Garufi2013,Fung2015,Bae2016,Stolker2016,Dong2017a,Casassus2021,Izquierdo_exoALMA}). While such companions were previously not found in a variety of high-contrast imaging programs (e.g. \citealp{Maire2017,Cugno2019,Zurlo2020,Asensio2021,Follette2023}), a potential candidate located at $\sim 300$\,au in the disk was identified by \cite{Cugno2024} in JWST data.   
\subsection*{HD\,143006}
\noindent The HD\,143006 system, located at a distance of $\sim 167\,\mathrm{pc}$ \citep {Gaia2023} in the Upper Scorpius star-forming region, consists of a G-type T\,Tauri star of age 7-12\,Myr \citep{Andrews2018,Garufi2018} and a disk seen close to face-on.

Imaging of the disk's infrared scattered light revealed various brightness asymmetries, including a broad dark region covering the west side of the disk, two narrow shadow lanes, and a prominent over-brightness located in the south-east of a ring at $\sim$ 18-30\,au \citep{Benisty2018}. A similar, yet less pronounced, over-brightness is also present in the outer disk ring at $\sim$ 50-83\,au. 

The mm-continuum on the other hand is marked by three concentric dust rings (8, 40, and 64\,au) and a bright crescent-like structure in the south-east, just outside the outer dust ring at 74\,au \citep{perez2018}. The latter is further substructured into three different peaks and generally consistent with a vortex formed at the edge of a cavity, with its morphology, however, pointing towards a second-generation vortex \citep{perez2018}. Moreover, the bright arc coincides with the outer over-brightness in the scattered light emission but is less extended, which matches predictions of dust trapping in vortices \citep{Baruteau2016}.

Observations of CO gas emission in HD\,143006 exhibit less structures but indicate gas depletion in the very inner disk and the presence of a kink-feature in some red-shifted channels \citep{perez2018,Pinte2020}. Both the scattered light and mm-observations support the scenario of a misaligned inner disk \citep{Benisty2018,perez2018}, with the morphology possibly being explained by the combination of a misaligned binary and an embedded aligned planet in the outer disk \citep{Ballabio2021}. 
\subsection*{HD\,34282}
\noindent The Herbig Ae star HD\,34282 -- or V1366 Ori -- is about 8.9\,Myr old \citep{Asensio2021} and located at a distance of $\sim$ 309\,pc \citep{Gaia2023}. A highly inclined circumstellar disk around the source was resolved in the mm-emission by \cite{Pietu2003}. A large inner dust cavity in the disk was suggested through modelling of the SED and mid-infrared Q-band images \citep{Acke2009,Khalafinejad2016} and later confirmed ($\sim$ 80\,au) in ALMA data by \cite{Plas2017}. These images further revealed that the $\sim$ 300\,au wide dust ring contains an azimuthal asymmetry in the south-east. The authors suggest a massive brown dwarf companion ($\sim$ 50\,M$_\mathrm{J}$) to be responsible for both observed substructures. 

Scattered-light observations allowed \cite{Boer2021} to identify two inclined rings and a tightly wound single-arm spiral, which is generally consistent with a Jupiter-mass planet and coincides with the mm-crescent. The latter may be the signature of vortex in the disk which, as shown by \cite{Marr2022}, can resemble a spiral arm in a highly inclined source. So far, no direct evidence of a companion in HD\,34282 has been found \citep{Asensio2021,Boehler2021,Quiroz2022}. The study of \cite{Bohn2022}, who compare position angle and inclination of the inner disc measured with VLTI/GRAVITY and the outer disc measured with ALMA, suggests a misalignment between the inner and outer disk of the system, yet no clear shadow signatures are visible in the scattered light images
\subsection*{MWC 758} 
\noindent Located in the Taurus star-forming region at a distance of 156\,pc \citep{Gaia2023}, the young ($\sim$ 8.9\,Myr; \citealp{Asensio2021}) Herbig Ae star MWC\,758 -- also known as HD\,36112 -- is surrounded by a low-inclination transition disk which has been the subject of a range of high resolution studies, revealing a complex asymmetric morphology. 

NIR scattered light observations unveiled two prominent spiral arms \citep{Grady2013,Benisty2015} and other non-axisymmetric arc-like features \citep{Benisty2015}. The detection of a third spiral arm and potential point-like source at $\sim$ 17\,au were reported by \cite{Reggiani2018}. Both gravitational instability \citep{Dong2015b} and massive planets \citep{Dong2015a,Bae2018,Baruteau2019,Calcino2020} have been suggested as the driving mechanism of the spirals, with the latter being favoured by an imaging study of the spiral pattern over 5\,yr \citep{Ren2020}. 

No fully depleted cavity was found in the scattered light data, but the mm-continuum of the disk is marked by a large and eccentric cavity ($\sim$ 40\,au; \citealp{Dong2018_mwc}), possibly carved by an embedded planet \citep{Isella2010,Andrews2011,Marino2015,Boehler2018}. Additionally, two bright emission clumps inside a double-ring structure are present at radii of $\sim$ 47 and 82\,au \citep{Marino2015,Boehler2018,Dong2018_mwc,Casassus2019b}, and may be related to dust-trapping pressure maxima in a vortex. 

Several spirals have been identified in both the continuum emission \citep{Boehler2018,Dong2018_mwc,Shen2020} and CO data \citep{Boehler2018}, which are partly co-located with the spirals observed in the infrared data. \cite{Baruteau2019} performed hydrodynamical simulations showing that the spirals, the eccentric and asymmetric ring structure, and the crescent-shaped features may all be explained by two Jupiter mass planets, one located inside and one outside the spirals. High-contrast imaging searches have reported the detection of such planets \citep{Reggiani2018,Wagner2019,Wagner2023}, but they remain to be confirmed.  
%
%
%
%
%
%
%
%
%
%
\section{Spiral features}\label{sec:spirals}
\begin{figure*}
\centering
\includegraphics[width=1.0\textwidth]{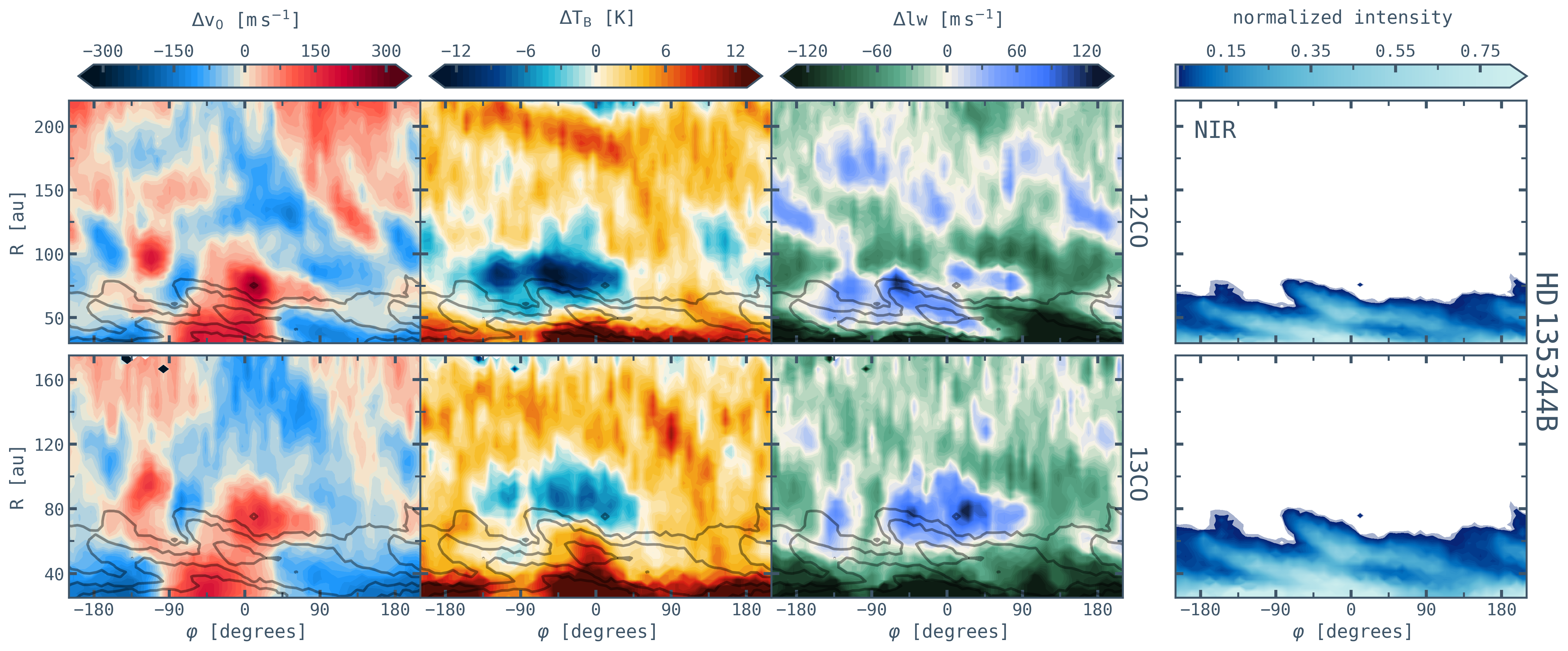}
\includegraphics[width=1.0\textwidth]{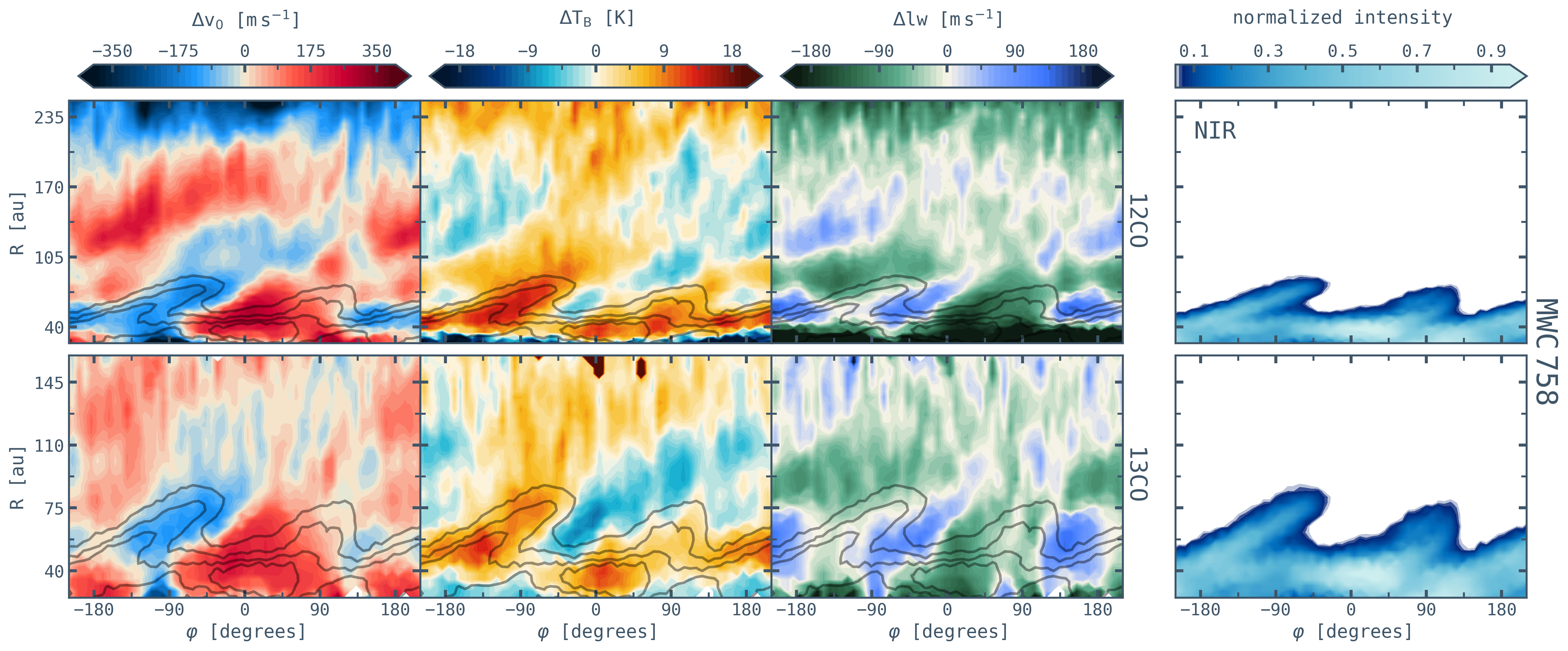}
\caption{Polar-deprojected residuals of the centroid velocity, peak intensity, and line width, shown for both CO lines and the HD\,135344B (top) and MWC\,758 (bottom) disks. The deprojection of the NIR scattered light emission is included in the forth panel of each row \citep{Stolker2017,Ren2023}, highlighting the observed spiral features which are overlaid as contours on the other three panels. The SPHERE images are normalized to the peak value and shown with a logarithmic normalization of the color map. Note the different spatial scale of the rows.}
\label{fig:polarSpirals}
\end{figure*}

\noindent In \autoref{fig:polarSpirals}, we present the polar-deprojected CO residuals for the centroid velocity, peak intensity, and line width alongside the polar-deprojected scattered light emission, for the two disks that show prominent spiral features: HD\,135344B and MWC\,758. For comparison, the spirals found in the near-infrared are further overlaid as contours on the gas emission residuals. The deprojection is performed by using the inclination and position angle (see \autoref{tab:dataInfo}, \citealp{Izquierdo_exoALMA}) of the individual sources. This is sufficient due to the face-on nature of the disk. 

For both disks, the inner spirals seen in the velocity residuals appear to align somewhat with the NIR spirals. In HD\,135344B, both NIR spiral arms are located just inside the red-shifted velocity arms. In MWC\,758, one of the NIR spiral ($-90$ to $90\degree$) has a red-shifted counterpart in the velocity residuals while the other has a blue-shifted counterpart ($-180$ to $-90\degree$). For the peak intensity, on the other hand, the spiral patterns seen for MWC\,758 align very well with the NIR spirals, whereas this is not the case for HD\,135344B. We note that the strong temperature decrement seen around 80\,au in this source is the result of continuum subtraction. In the line width, spiral patterns are mostly visible in the negative residuals, which again seem to somewhat align with the NIR spirals for MWC\,758 but not HD\,135344B (for a detailed discussion see Benisty et al. in prep.). 
%
%
\section{Unconvolved model images}\label{sec:model_unconv}
\begin{figure*}
\centering
\includegraphics[width=1.0\textwidth]{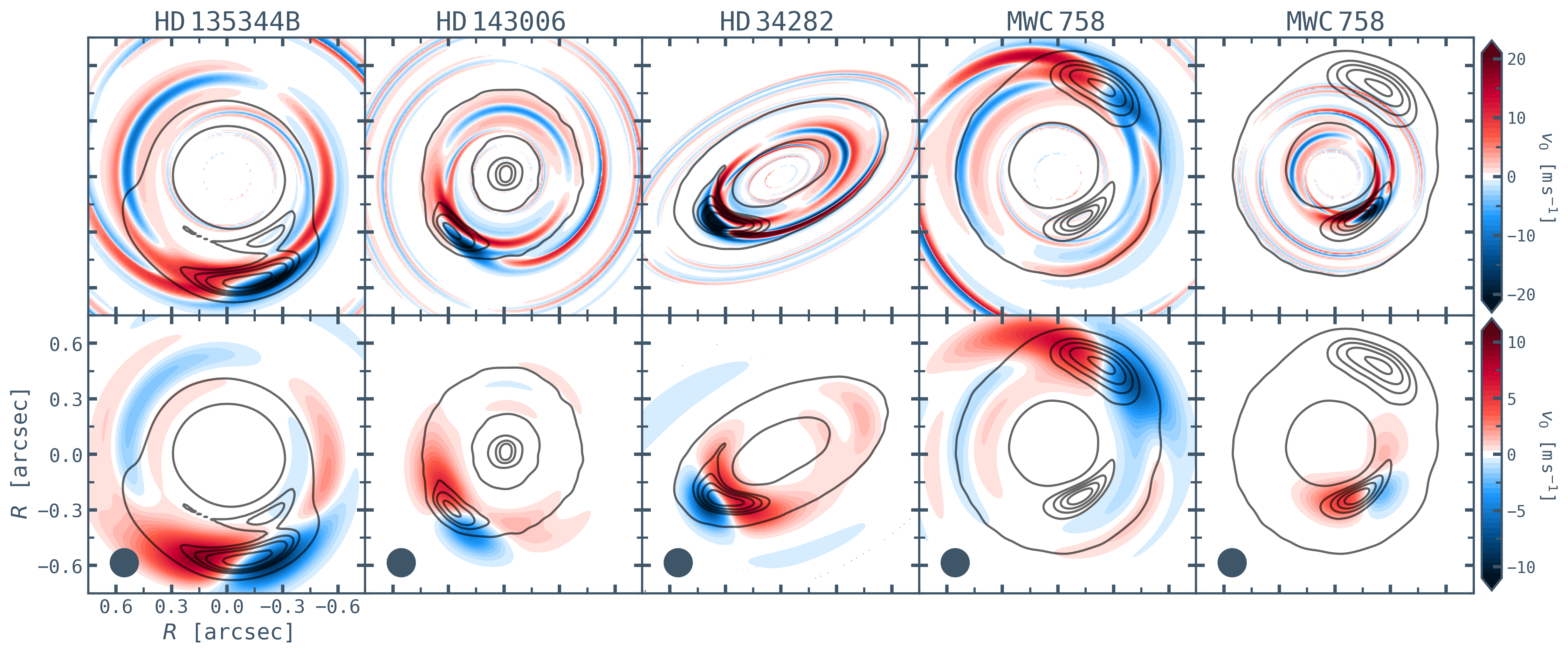}
\caption{Velocity perturbations of the dead zone vortex simulation, post-processed with RADMC-3D and adapted to each dust asymmetry observed in the data. Shown are the unconvolved models (top row), and models convolved with the beam of the observation (bottom row). The continuum observations are overlaid in all panels to highlight the location of the dust asymmetries.}
\label{fig:vortexCompareUnconv}
\end{figure*}

\noindent In \autoref{fig:vortexCompareUnconv}, we show the unconvolved model images of the dead zone simulation alongside the convolved images (0\farcsec15 beam). The continuum observations are overlaid as contours to highlight that the model has been adapted to each disk such that the vortex is present at the same location as the dust asymmetries. In this work, we are mainly interested in the kinematical patterns of a vortex and we note that the exact strength of these features depends on various factors. Moreover, we assume a flat disk and the magnitude of the vortex may be significantly underestimated. In that regard, the unconvolved images nevertheless emphasize the need for higher spatial resolution. In particular, an inclined source such as HD\,34282 may show very strong vortex features if resolved at very high resolution.     
%
%

\bibliography{bibliography}{}
\bibliographystyle{aasjournal}

\end{document}